\documentclass[
aps, prl,
twocolumn, 
showpacs, 
nopreprintnumbers, 
amsmath,amssymb,
superscriptaddress,
letterpaper
]{revtex4}

\usepackage{graphicx}
\usepackage{dcolumn}
\usepackage{bm}

\begin{document}


\title{ 
Decoherence and localization in tunneling process under influence of one
external degree of freedom
}

\author{Quanlin \surname{Jie}}
\email[E-mail: ]{qljie@whu.edu.cn}
\affiliation{%
Department of Physics, Wuhan University,
Wuhan 430072, P. R. China
}%

\author{Bambi \surname{Hu}}
\affiliation{%
Department of Physics and Centre for Nonlinear Studies, Hong
Kong Baptist University, Hong Kong, China
}
\affiliation{%
Department of Physics, University of Houston, Houston TX 77204-5506
}

\author{Guangjong \surname{Dong}}
\affiliation{%
Department of Physics and Centre for Nonlinear Studies, Hong
Kong Baptist University, Hong Kong, China
}
\date{\today}

\begin{abstract} 
We investigate numerically the tunneling effect under influence of another
particle in a double well system. Such influence from only one degree of
freedom makes decoherence and quantum-classical transition, i.e., suppression
of the tunneling effect. The decoherence happens even for cases that the
influence is from a particle of very small mass, and it has virtually no
effect in the corresponding classical dynamics. There are cases similar to
dynamical localization that the suppressed tunneling rate is several times
smaller than the classical counterpart. This result is relevant for
understanding quantitatively the dynamical process of decoherence and quantum
to classical transition.
\end{abstract}

\pacs{03.65.Yz, 03.65.Xp, 05.45.Mt, 03.65.Sq}
\keywords{Tunneling effect, Decoherence, Quantum-classical transition}
\maketitle

%


Decoherence is a time honored problem that attracts many investigations for
the mechanism of quantum to classical transition~\cite{1,2}. There are growing
interests in decoherence in the field of quantum information~\cite{3,4,5}. A
common approach to decoherence is to study systems that thermally interact
with large degrees of freedom, such as enviroment. This approach needs
statistical method to handle the large degrees of freedom of the enviroment,
and one usually resorts to master equation~\cite{6,7,8} or semiclassical
approach~\cite{9} for the reduced density matrix. There is little
investigation of decoherence that involves few degrees of freedom. One usually
expects quantum effect prominent when there are only a few degrees of freedom
involved in a physical process. However, as shown in Ref.~\cite{10}, if a time
dependent system is classical chaotic, such as kicked rotator, decoherence
happens under influence of its one internal degree of freedom. The chaotic
nature of the underlying dynamics makes the internal degree of freedom to
behave as a noise~\cite{10a,10b}.




In this Letter, we investigate numerically the tunneling process of a particle
in a one dimensional double well system under influence of another particle.
Tunneling is a basic quantum effect that has far reaching applications.
Understanding decoherence in tunneling effect is of fundamental
importance~\cite{15,16,17}. Our numerical results show that the tunneling can
be totally suppressed under influence of only one external degree of
freedom. Even such influence has virtually no effect in the corresponding
classical dynamics, it is still able to cause decoherence and hence makes the
tunneling rate to approach the classical counterpart. Another result is that,
similar to dynamical localization, the suppressed tunneling rates are several
times smaller than the classical counter part in some cases.




The Hamiltonian of the two-particle system reads
$H=\sum_{i=1,2}\left[\frac{p_i^2}{2m_i}+v_i(x_i)\right]+v(x_1-x_2).$ Here we
consider the motion of particle $1$ under influence of particle $2$ through
the interaction $v(x_1-x_2)$. Our tests indicate that the results are rather
insensitive to the exact form of the interaction provided that the total
system is not integrable classically. We use an attractive interaction
$v(x_1-x_2)=\frac12\gamma(x_1-x_2-l_0)^2$ in the following discussions. The
double well potentials for both particles are a Gaussian shaped barrier at the
bottom of a bounded harmonic potential, $v_i(x_i)=\frac12k_ix_i^2+\lambda_i
\exp(-x_i^2/a_i^2)$. This form of double well performs similar to the common
used form $\lambda(x^2-a^2)^2$. Yet it is more flexible to control the height
and width of the barrier by adjusting parameters $\lambda_i$ and $a_i$, and
hence is more relevant for experimental realization by imposing a barrier at
the minimum of a parabolic trap~\cite{18}.

We consider each of the two particles initially as a Gaussian wave packet
centered at the bottom of one well with vanished central momentum,
$\phi_0(x_i)=A\exp(-\alpha (x_i-x_{0i})^2/2\hbar)$, where $A$ is a
normalization constant, $\alpha$ is the width parameter, and $x_{0i}$ is
center of the wave packet. When the interaction $v(x_1-x_2)$ vanishes, the
behavior of the particle $1$ is well known~\cite{11}.
%
%
The quantum motion has virtually only finite number of frequencies, i.e., the
initial wave packet is effectively spanned by only finite number of
eigenfunctions of the Hamiltonian. In our model, only $2$ frequencies
actually dominate the one dimensional tunneling processes. 




\begin{figure}[h]
\includegraphics[angle=-90,width=0.8\columnwidth,clip]{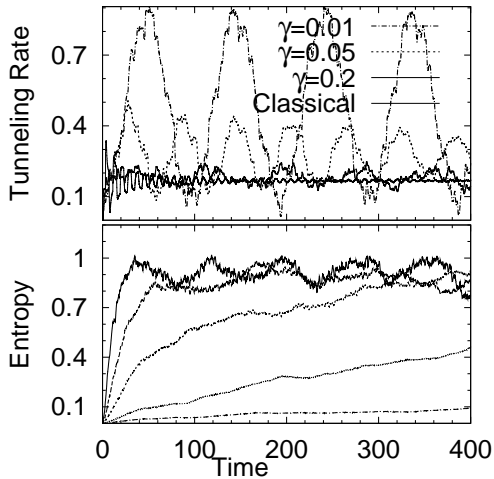}
\caption{\small
(Top) Tunneling rates versus time for interaction strength $\gamma=$ $0.01$
(dot-dashed),$0.05$ (dashed), and $0.2$ (thick solid), respectively. The thin
solid line is the classical counterparts. (Bottom) The von Neumann entropy
versus time. From bottom to top, $\gamma=0.01$, $0.02$, $0.05$, $0.1$, $0.2$.
}\label{fig1}
\end{figure}


When the particle is under influence of other degrees of freedom, the
tunneling behavior may changes completely. From our numerical tests, the
suppression of the tunnel effect happens under influence of only one degree of
freedom, and such influence is classically negligibly small.  In
Fig. \ref{fig1}, we show the tunnel effect of particle $1$ under influence of
particle $2$ with a mass that is $10^4$ times smaller than that of particle
$1$, $m_2=10^{-4}m_1$.  The top of Fig. \ref{fig1} is the tunneling rates of
particle $1$ versus time for various interaction strength $\gamma$. The
particle $1$ with unit mass is initially rest on the bottom of right well. The
tunneling rate (also known as tunneling probability) is the probability to
find the particle on the other side of the barrier,
$T_r=\int_{-\infty}^0 \rho_1(x_1)dx_1.$
Here $\rho_1=\int |\psi(x_1,x_2)|^2 dx_2$ is the reduced density for the
particle $1$ by tracing out the other degree of freedom from wave function
$\psi(x_1,x_2)$. The double well potential parameters (in arbitrary unit) are
$k_1=0.5$, $\lambda_1=3.0$, $a_1=1.0$, and $l_0=0.5$ (The Planck constant
$\hbar$ is set to $1$). The potential for the particle $2$ is the same as that
of particle $1$. We set particle $2$ to be the same Gaussian wave packet as
particle $1$ that initially locates in the bottom of another well with
vanished central momentum. The width parameter for both Gaussian wave packets
is $\alpha=3.0$. In classical limit, the particle $2$ with such a small mass
has virtually no effect on the motion of particle $1$. The quantum case,
however, is totally a different story. The interaction with particle $2$ can
completely destroy coherence of particle $1$. As shown in Fig. \ref{fig1}, for
very weak interaction, $\gamma=0.01$, the motion of particle $1$ is almost
unaffected. It is still able to tunnel through the barrier back an forth
almost completely. When the interaction becomes a little bit stronger,
$\gamma=0.05$, the decoherence is evident. The maximum tunneling rate is at
most $50\%$, and such maximum decreases gradually in longer time. Further
increase of the interaction strength to $\gamma=0.2$, as shown by the thick
solid line, the tunneling effect is virtually suppressed by the
decoherence. The resultant tunneling rate approaches to the classical
result. As a comparison, we show the corresponding classical counterpart by
thin solid line. It is obtained by evolving the Liouville equation from a
classical density distribution that is identical to the quantum Wigner
function initially. Similarly, the classical counterpart of the tunneling rate
is the probability to find the particle on the other side of the barrier.



Since the mass of particle $2$ is smaller than particle $1$ by several orders,
its effect on particle $1$ is just like a small weightless dust on a heavy
body in the classical limit. The classical motion of particle $1$ is indeed
quite regular, and is almost the same as the unperturbed one dimensional
case. The classical motion of particle $2$, however, is irregular. The
interaction with particle $1$ is a major force for the chaotic motion of
particle $2$ which bounces drastically between the two wells. Such
irregularity of particle $2$ makes the total wave function of the two-particle
system to be irregular. Entanglement between the two particles leads to the
decoherence of the quantum motion of particle $1$. The irregular motion of
particle $2$ poses some difficulties in numerical calculations. The wave
function relates to particle $2$ spreads widely in both coordinate and
momentum space. One needs dense grid that covers a wide range. Our actual
calculations employ periodic boundary condition to particle $2$. This indeed
restricts particle $2$ to a ring. From our test, this restriction does not
alter the behavior of particle $1$, but saves computation efforts remarkably.



There are more frequencies involved in the two-particle system, and the
structure of the frequencies is much complicated than the single particle
case. One can obtain the spectrum density by Fourier transformation of the
auto-correlation function. Since the mass of particle $2$ is much smaller than
that of particle $1$, the energy spectrum distributes in a wide range with
band like structure. Each band consists of several levels. It is virtually
impossible to have constructive interference between those frequencies on the
other side of the barrier.



The complex structure of the involved frequencies and the corresponding
eigenfunctions relates to the entanglement of between the two particles. Such
entanglement makes expansion of initial wave packet in terms of the
eigenfunction to be complicated. An initially wave packet must evolve into
entangled form in such system.  The bottom of Fig. \ref{fig1} shows the von
Neumann entropy of particle $1$ versus time for various interaction
strengths. The von Neumann entropy, $S=Tr(\rho_1 \ln(\rho_1))$, is a common
used measure for the entanglement between two particles. By comparison with
top part of Fig. \ref{fig1} for the correspondent interaction strength, it is
evident that the decoherence in the tunneling process corresponds to the
entanglement between the two particles. Increase the interaction between the
two particles leads to stronger entanglement (larger entropy), and hence
deeper suppression of the quantum effects. An initially Gaussian wave packet
needs some time to develop into the final entangled form. The speed of such
development increases with the interaction strength.


\begin{figure}[h]
\includegraphics[angle=-90,width=0.8\columnwidth,clip]{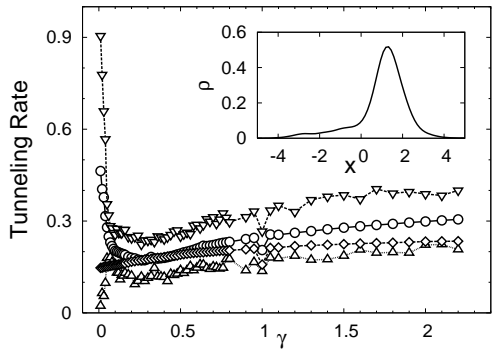}
\caption{\small
The tunneling rate versus interaction strength $\gamma$.  The circle
(diamond), down and up triangle symbols correspond to quantum (classical) mean
value, typical maximum and minimum value of the quantum tunneling rates,
respectively. (Inset) The density distribution $\rho$ of particle 1 versus the
coordinate $X$ at time $t=400$ for $\gamma=0.2$.
}\label{fig2}
\end{figure}

When the interaction is too strong, the two particles can bind together like a
single one. When this happen, the quantum coherence emerges again. In other
words, when the interaction strength is stronger than some value, the degree
of freedom for the relative motion is hard to excite, and there is enhancement
of the quantum coherence with the increase of the parameter $\gamma$.  Note
that when two particles bind together, they behave like a particle of mass
$m_1+m_2$ in a potential $v_1+v_2$. Here $m_1$, $m_2$ and $v_1$, $v_2$ are the
masses and potentials of particles $1$ and $2$, respectively. Fig. \ref{fig2}
shows averaged tunneling rate of particle $1$ versus the interaction parameter
$\gamma$ for both quantum and the corresponding classical cases. The open
circles and diamonds are quantum and classical mean tunneling rate,
respectively. The down and up triangles are a pair of typical maximum and
minimum tunneling rates chosen from a tunneling cycle. The range between such
a pair of the triangles relates to the degree of coherence in the tunneling
process. The initial states of the two particles and potential parameters are
the same as that of Fig. \ref{fig1}. In classical limit, because of small
mass, the particle $2$ possesses little energy and momentum to affect the
motion of particle $1$. In contrast, the quantum motion of particle $1$
exhibits decoherence by interaction with particle $2$ for a wide range of
interaction strength. As the interaction strength parameter $\gamma$ grows
from $0$, the averaged quantum tunneling rate approaches to classical limit
exponentially. Then quantum and classical mean tunneling rates coincide for a
wide range of strength parameter $\gamma$. When $\gamma>0.8$, the binding
energy between the two particles becomes hard to excite, quantum tunneling
rate increases and deviates from the classical counterpart. Further increase
of $\gamma$ makes the quantum motion to gain more coherence, and to behave
more like the motion of a single particle. In all the cases, corresponding to
the regular classical motion, the reduced density distribution for particle
$1$, as shown in the inset of Fig. \ref{fig2}, is quite regular. Another point
to note is that when the quantum mean tunneling rate approaches to the
classical counterpart, its fluctuation, as shown by the range between up and
down triangles, is larger than the classical case. This is resultant from the
fact that the dynamics of the two-particle system has only finite number of
frequencies.


\begin{figure}[h]
\includegraphics[angle=-90,width=0.8\columnwidth,clip]{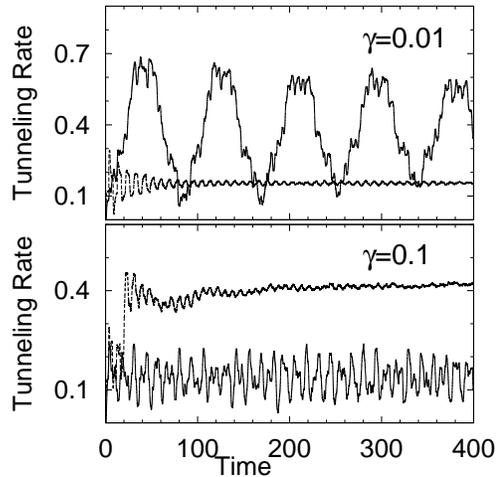}
\caption{\small
Tunneling rate versus time for interaction strength $\gamma=0.01$ (Top), and
$\gamma=0.1$ (Bottom), respectively. The solid and dashed lines are quantum
and classical results, respectively.
}\label{fig3}
\end{figure}


In some cases, the fact of finite frequencies makes the quantum tunneling rate
several times smaller than the classical counterpart. Such cases occur at some
interaction strengths and the masses of the two particles close to each
other. We show such a case in Fig \ref{fig3}. The parameters for the on site
potentials and interaction, as well as the initial states of the two
particles, are the same as that of Fig. \ref{fig1} except $\lambda_2=15$. The
masses of the two particles are same, $m_1=m_2=1$. As shown in the top part of
Fig. \ref{fig3}, when the interaction is weak ($\gamma=0.01$), the quantum
coherence is almost untouched. As the strength of interaction increase,
quantum and classical behaviors change in opposite way. Quantum tunneling rate
is quickly suppressed by decoherence. The classical counterpart, however,
increases with the interaction strength. This is because the two particles are
initially located on the different sides of the barrier, each particle has a
pull force on the other one. As the parameter $\gamma$ increase, they pull
each other with larger force. This leads to increase on the classical motion's
energy to climb over the barrier. When the parameter $\gamma=0.1$, as shown in
the bottom of Fig. \ref{fig3}, the quantum tunneling rate is about $3$ times
smaller than its classical counterpart.


This effect, similar to the dynamical localization~\cite{21}, results from
destructive interferences between involved frequencies. It relates to two
factors: decoherence from interaction with particle $2$ and only finite
frequencies involved in the quantum motion. In classical limit, the attractive
interaction between the two particles makes each particle gaining energy from
other one for climbing over the barrier. The energy transformation between the
two particles is prominent when the frequency of the relative motion matches
the overall motion, i.e., resonance between relative motion and overall
motion. For the given parameters, the system is close to the classical
resonance when $\gamma=0.1$. For quantum motion, however, both relative and
overall motions contain only finite frequencies. There is little chance that
the two kinds of frequencies match. In other words, the quantum motion doesn't
feel the resonance. On the other hand, the interaction with particle $2$
destroys coherence between these involved frequencies. It is virtually
impossible to make constructive interference on the other side of the
barrier. Thus the quantum motion behaves like a classical motion without the
resonance effect.  This is in fact a kind of quantum dynamical localization to
the classical resonance.

\begin{figure}[h]
\includegraphics[angle=-90,width=0.8\columnwidth,clip]{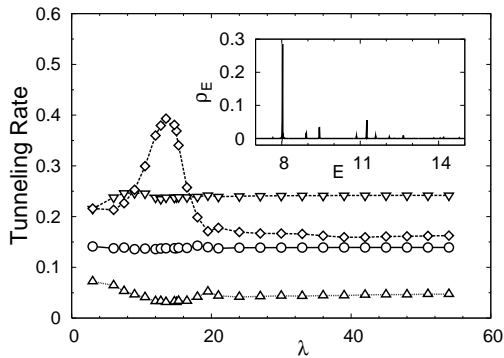}
\caption{\small
Tunneling rate versus the barrier height $\lambda$ of the particle $2$. The
circle (diamond), down and up triangle symbols have the same meaning as that
of Fig. \ref{fig2}. (Inset) Density spectrum $\rho_E$ versus energy $E$ for
$\lambda=15$.
}\label{fig4}
\end{figure}

To verify the above explanation, we rerun the calculation with different
barrier hight for particle $2$. The result is shown in Fig. \ref{fig4}. The
symbols and their meanings are the same as that of Fig. \ref{fig2} except the
horizontal axis is the barrier's height of particle $2$. As the height of the
barrier increase, the particle $2$ becomes harder and harder to move across
barrier, and it is virtually trapped in the left well. This makes the
classical motion alway from the resonance region. When this happens, as shown
in Fig. \ref{fig4}, the classical tunneling rate drops. The restriction on the
particle $2$ has virtually no effect on the quantum motion of particle
$1$. Thus when the classical dynamics leaves the resonance region, the quantum
tunnel rate approaches to its classical counterpart. As shown by the inset,
when the masses of the two particles close to each other, there are less
involved frequencies than the cases shown in Fig. \ref{fig1} and
Fig. \ref{fig2}. This means the involved frequencies have larger level space,
i.e., the frequencies are more discrete. 





In conclusion, we numerically show that decoherence and quantum-classical
transition occur in the tunneling process under influence of only one
external degree of freedom. Such influence can be negligibly small in the
classical limit. Here, the correspondent classical motion of the one
dimensional tunneling is regular when there is no interaction with external
degree of freedom, and such regularity of classical motion is almost unchanged
by such negligible interaction. However, the interaction makes the motion of
external degree of freedom to be chaotic. It is such irregularity of the
external degree of freedom that causes decoherence in the quantum motion by
entanglement with the external degree of freedom. Since there is only finite
number of frequencies involved in quantum motion, the quantum motion is unable
to feel classical resonance. This makes quantum tunneling to behave similar to
the dynamical localization in the kicked rotator, i.e., the tunneling rate is
several times smaller than the classical counterpart in some cases.



\bigskip

This work is supported in part by the National Science Foundation
(Grant No. 10375042), grants from the Hong Kong Grants Council (GRC),
and the Hong Kong Baptist University Faculty Research Grant (FRG).


\end{document}